\documentclass[usenatbib,referee]{mn2e}
\input psfig.sty
\usepackage{epsf}
\usepackage{graphicx}

\def\eps@scaling{1.0}%
\newcommand\epsscale[1]{\gdef\eps@scaling{#1}}%
\newcommand\plotone[1]{%
 \centering
 \leavevmode
 \includegraphics[width={\eps@scaling\columnwidth}]{#1}%
}%

\def\apj{ApJ}
\def\mnras{MNRAS}
\def\apjs{ApJS}

\def\beq{\begin{equation}}
\def\eeq{\end{equation}}
\def\bey{\begin{eqnarray}}
\def\eey{\end{eqnarray}}
\def\pppm{\rm P^3M}
\def\Mpc{\,{\rm Mpc}}

\def\mpchi{\,h^{-1}{\rm {Mpc}}}

\def\kpchi{\,h^{-1}{\rm {kpc}}}
\def\kms{\,{\rm {km\, s^{-1}}}}
\def\msun{{M_\odot}}
\def\msunhi{\,h^{-1}{M_\odot}}
\def\Rvir{R_{\rm vir}}
\def\vn{v_{\rm n}}
\def\vt{v_{\rm t}}

\def\thetaspin{\theta_{\rm spin}}

\def\gs{\mathrel{\raise1.16pt\hbox{$>$}\kern-7.0pt
\lower3.06pt\hbox{{$\scriptstyle \sim$}}}}
\def\ls{\mathrel{\raise1.16pt\hbox{$<$}\kern-7.0pt
\lower3.06pt\hbox{{$\scriptstyle \sim$}}}}
\def\gtsima{$\; \buildrel > \over \sim \;$}
\def\ltsima{$\; \buildrel < \over \sim \;$}
\def\prosima{$\; \buildrel \propto \over \sim \;$}
\def\gsim{\lower.5ex\hbox{\gtsima}}
\def\lsim{\lower.5ex\hbox{\ltsima}}
\def\simgt{\lower.5ex\hbox{\gtsima}}
\def\simlt{\lower.5ex\hbox{\ltsima}}
\def\simpr{\lower.5ex\hbox{\prosima}}
\def\la{\lsim}
\def\ga{\gsim}

\def\Nsub{N_{\rm sub}}
\def\Nhalo{N_{\rm halo}}
\def\Msub{M_{\rm sub}}
\def\Mhalo{M_{\rm host}}
\def\tcr{t_{\rm cr}}
\def\vvir{v_{\rm vir}}
\def\rvir{r_{\rm vir}}
\def\Rvir{R_{\rm vir}}
\def\thetaInfall{\alpha_{\rm infall}}

\title
[The phase-space distribution of infalling dark matter subhalos]
{The phase-space distribution of infalling dark matter subhalos}
\author[Wang, Jing, Mao \& Kang]{
H.Y. Wang$^{1,2}$
\thanks{whywang@mail.ustc.ac.cn, ypjing@shao.ac.cn, smao@jb.man.ac.uk,
        kangx@shao.ac.cn},
Y.P. Jing$^{2}$, Shude Mao$^3$, Xi Kang$^2$ \\
${^1}$Center for Astrophysics, University of Science and Technology
      of China, Hefei, Anhui 230026, China \\
${^2}$Shanghai Astronomical Observatory, the Partner Group of
MPI f\"ur Astrophysik, Nandan Road 80, Shanghai 200030, China \\
${^3}$University of Manchester, Jodrell Bank Observatory,
  Macclesfield, Cheshire SK11 9DL, UK}

\date{
Accepted ........
Received .......;
in original form ......}

\pubyear{2005}
\begin{document}
\maketitle

\begin{abstract}
We use high-resolution numerical simulations to study the physical
properties of subhalos when they merge into their host halos. An
improved algorithm is used to identify the subhalos. We then
examine their spatial and velocity distributions in spherical and
triaxial halo models. We find that the accretion of
satellites preferentially occurs along the major axis and
perpendicular to the spin axis of the host halo. Furthermore, the
massive subhalos show a stronger preference to be accreted along
the major axis of the host halo than the low-mass ones.
Approximate fitting formulae are provided for the physical
properties of subhalos. Combined with analytical and semi-analytic
techniques, these empirical formulae provide a useful basis for
studying the subsequent evolution of subhalos and satellite
galaxies in their hosts. Future studies should however
account for satellites that may not be undergoing the first infall
in their evolution.
\end{abstract}


\section {Introduction}
Cold Dark Matter (CDM) dominated cosmological models have received
strong support from a wide range of observations (e.g., Spergel et
al. 2003). In CDM models, the structures of dark matter are formed
through the merger and accretion of smaller structures (White \& Rees
1978).  Dark matter halos are high-density structures that are a few
hundred times of the mean background density and are in approximate
dynamical equilibrium according to the virial theorem.  Galaxies are
expected to condense in dark matter halos due to dissipative cooling
processes.  The observable properties of the galaxies are thus
strongly influenced by the hierarchical growth of their host halos.
During the hierarchical assembly of the halos, smaller halos merged
into bigger ones are found to be long-lived even though their outer
part may be stripped by the tidal interactions (e.g., Tormen 1997;
Klypin et al. 1999; Moore et al. 1999).  Some of these subhalos are
believed to be the hosts of the satellite galaxies in galactic-sized halos
(such as the Milky Way)
and of the galaxies in clusters of galaxies, therefore it is essential
to understand the evolution and dynamics of these subhalos in galaxy
formation. Indeed, it has been demonstrated that it is important to
resolve the merger of subhalos in order to accurately predict the
cluster luminosity function of galaxies (Springel et al. 2001) and the
luminosity function of field galaxies (Kang et al. 2004). The presence
of the subhalos was used to interpret the anomalous flux ratios of
lensed quasar images (e.g., Mao \& Schneider 1998; Kochanek \& Dalal
2004), though it is still unclear if the amount of subhalos predicted
by the CDM models is in quantitative agreement with the lensing
observations (Mao 2004; Mao et al. 2004; Zentner et al. 2005a).
In addition, the subhalos
have important effects on the heating of galactic disks (Benson et
al. 2004) and possible $\gamma$-ray emission from annihilation of
dark matter particles (Taylor \& Silk 2003; Stoehr et al. 2003; see Bertone et
al. 2004 for a review.)

Because of the importance of the subhalos in cosmological studies,
there have been many high-resolution N-body studies of subhalos since
Moore et al. (1999) who discovered that subhalos can survive
within bigger halos for a significant period of time. The mass
function of subhalos was shown to be proportional to $\sim
(\Msub/\Mhalo)^{-1.8}$ (e.g., Klypin et al. 1999; Moore et al. 1999;
Ghigna et al. 2000), and the radial distribution of subhalos above a
given mass is much flatter than that of the dark matter in the host
halo (e.g., Gao et al. 2004a; Diemand et al. 2004; Mao et al. 2004).
The velocity dispersion of the subhalos is larger than that of the
underlying dark matter (positive velocity bias), except in the
innermost region where the velocity anti-bias is found for the
subhalos (Ghigna et al. 1998; Okamoto \& Habe 1999; Colin et al. 2000;
Klypin et al. 1999; Springel et al. 2001; Diemand et al. 2004).  In
their recent work, Gao et al. (2004a) showed that most present-day
subhalos identified in their simulations have been accreted into their
host halos very recently. This indicates that the
subhalos accreted earlier, even if they were massive enough to host bright
galaxies, may have been disrupted in the current generation of
high-resolution simulations (Gao et al. 2004b) due to physical (such
as tidal stripping) and numerical effects (e.g., limited resolution). It is
therefore very demanding resolution-wise to correctly model the dynamics,
mergers and evolution of subhalos and satellites in the densest regions (Kang et
al. 2004), and thus an analytical theory would be preferred
to follow the evolution of subhalos within their hosts.

Indeed, parallel to the numerical studies, significant progress has
been made in understanding the evolution and the distribution of
subhalos with analytical models (e.g., van den Bosch et al. 1999;
Klypin et al. 1999; Taylor \& Babul 2002, 2004; Sheth 2003;
Zentner \& Bullock 2003; Ougri \& Lee 2004;
van den Bosch, Tormen \& Giocoli 2005). The two principal physical
processes, dynamical friction and tidal stripping, which determine the
evolution of subhalos, are relatively well understood. However, as
noted by Benson (2005), the initial conditions (velocities and positions)
of the infalling subhalos adopted in these studies are not from the
prediction of hierarchical CDM models.  Although some of these studies
predicted subhalos with physical properties in reasonable agreement
with those found in simulations, their success depends in part on the
fine-tuning of the model parameters, such as the Coulomb logarithm
function in the dynamical friction formula (e.g., Oguri \& Lee
2004). Obviously, in order to make such an analytical approach
more useful for studying the evolution of subhalos and satellite
galaxies, it is a prerequisite to use the correct
initial phase space information of the infalling subhalos
as found in the CDM cosmological model.

Several authors have already studied the orbital parameters of the
subhalos at the time of their mergers with the host halo (Tormen 1997;
Vitvitska et al. 2002; Khochfar \& Burkert 2004; Benson 2005). Tormen
(1997) and Khochfar \& Burkert (2004) used high-resolution
re-simulations of halos and identified the progenitors of these
halos. The orbital distribution is then measured from the progenitors
that are about to merge with the main progenitor.  In contrast,
Vitvitska et al. (2002) and Benson (2005) identified pairs of halos
that are about to merge, and measured the orbital distribution of
these pairs. Among these studies, Benson (2005) used a large set of
cosmological simulations provided by the Virgo Consortium (Jenkins et
al. 2001), and formed a large sample of such orbital pairs. From this,
he presented fitting formulae for the distribution of the initial
infall velocity of the subhalos at the virial radius of their host
halo.

Although the work of Benson (2005) is an important step forward to
determine the initial condition of infall subhalos, there is an
important question yet to be examined quantitatively,
i.e., whether the mergers are isotropic in the position space or
there is some preferential direction for the mergers. Earlier
studies suggest the accretion is anisotropic (e.g., Tormen 1997).
For example, Aubert, Pichon \& Colombi 2004) found the infall
takes place preferentially  in the plane perpendicular to the
direction defined by the spin of the halo. In this paper, we use
a cosmological N-body simulation of $512^3$ particles (Jing \&
Suto 2002; Kang et al. 2004) to investigate this problem
in more detail. As we will show, the mergers are preferentially
along the major axis of the host halos; we will quantify
this anisotropic accretion as a function of the subhalo mass (see
also Libeskind et al. 2005). Compared with the simulations used
by Benson (2005), our simulation has higher force and mass
resolutions, and the subhalos are better resolved. This makes it
easier and more reliable to quantify the orbital distribution of
subhalos. This is particularly important for small subhalos as
they may lose their identities in low resolution N-body
simulations when they are close to to the boundary of bigger
halos. It is difficult to account for this population of missing
subhalos if only one snapshot of the simulation output is used.

The paper is structured as follows. In \S2, we briefly describe the numerical
simulations we use, and how the subhalos are identified. In \S 3,
we study the physical properties of subhalos and present simple
empirical fitting formulae to these properties. In \S 4 we
summarize our main results and discuss areas for future improvement.

\section {N-body simulation}

The simulation used in this paper is a $\pppm$ cosmological simulation
of $512^3$ particles in a box of $100\mpchi$. The cosmological model
is the standard concordance model with the density parameter
$\Omega_{\rm m,0}=0.3$, the cosmological constant
$\Omega_{\Lambda,0}=0.7$ and the Hubble constant
$h=H_0/(100\,{\kms\Mpc^{-1}})=0.7$.  The initial density field is
assumed to be Gaussian with a Harrison-Zel'dovich primordial power
spectrum and with an amplitude specified by $\sigma_8=0.9$, where
$\sigma_8$ is the r.m.s. fluctuation of the linearly evolved density
field in a sphere of radius $8\mpchi$.  This simulation, which started
at redshift $z_i=72$, is evolved by 5000 time steps to the present day
($z=0$) with our vectorized parallel $\pppm$ code (Jing \& Suto
2002). The force softening length $\eta_f$ (S2 type, Hockney \&
Eastwood 1981) is $10\kpchi$ comoving, and the particle mass
$m_p=6.2\times 10^{8}\msunhi$. Because these simulation parameters are
very similar to those adopted in many high-resolution re-simulations
of individual cluster halos (Moore et al. 1999, Jing \& Suto 2000,
Fukushige \& Makino 2001, 2003, Power et al. 2003, Diemand et al
2004), we have achieved a resolution that can resolve subhalos within
massive host halos.

The dark matter halos are identified  at redshift $z=0$ in the
simulation described above using the
Friends-of-Friends method (FOF) with a linking length equal to 0.2
of the mean particle separation.  The subhalos are then identified
within the FOF halos with the {\tt SUBFIND} routine (Springel et
al. 2001). In Kang et al. (2004), the mass function of the
subhalos was examined, and was found to be in good agreement with
the subhalo mass functions obtained in previous halo
re-simulations (Springel et al. 2001) down to a mass of about
$3.1\times 10^{10} \msunhi$ (50 particles). This indicates that
the subhalos with more than $\Nsub=50$ particles are resolved in
the simulation. Considering the fact that subhalos can survive
more easily in the outskirt than in the inner dense region of a
host halo, we relax the lower limit to 20 particles for the
subhalo mass. Table \ref{table:haloSelection} lists the number of
particles we use to identify the subhalos and halos in several
different combinations. And we only use the simulation output at
z$=0$ to calculate the distribution.

\section{Results}

\subsection{The phase-space distribution in the spherical halo model}

We first consider the density profile of host halos as a
sphere and search for subhalos within a spherical shell with
radius $r$ between $1-\Delta r$ and $1+\Delta r$,  where
$r$ is the distance between the centres of the subhalo and the
host halo in units of the virial radius $\rvir$ of the host halo.
We identified the ``centres" of the halos as the lowest-potential
particles using the {\tt SUBFIND} routine (Springel et al. 2001;
Kang et al. 2004); we also checked that the ``centre" identified
this way is very close to the centre of mass for most host halos.
We normally take the thickness of the shell ($\Delta r$) to be
0.1 or 0.2 (see Table \ref{table:haloSelection}). The virial
radius is determined according to the spherical collapse model
(Kitayama \& Suto 1996; Bryan \& Norman 1999). In this section, we
consider only those subhalos with an inward directed velocity. We
treat the subhalos and host halos as point-mass particles, and
determine the velocity of the subhalos at the time when their
orbits first cross the virial radius of the host halo under
gravity. We also compute the time $\tcr$ for each subhalo to cross
from $1+\Delta r$ to $1-\Delta r$ according to their trajectories.
Subhalos with a larger $\tcr$ will stay longer in the shell of
$1\pm\Delta r$ than those with smaller $\tcr$. During a time
interval $dt$, the chance of observing this subhalo within the
spherical shell will be $dt/\tcr$. Thus we weight each subhalo by
$\tcr^{-1}$ when we compute the distributions of position, radial
and tangential velocities. We will excise the subhalos that do not
pass through one or both of the radial limits.

The distributions of the radial ($v_r$) and tangential
($v_\theta$) velocities of these subhalos  are presented in Fig.
1; both velocities are in units of the circular velocity at the
virial radius, $\vvir$. For this exercise, we only retain host
halos with more than 500 particles and subhalos with more than 20
particles, corresponding to the selection parameter set A in Table
\ref{table:haloSelection}.  The $v_r$ distribution peaks around
the virial velocity, while $v_\theta$ peaks at a slightly smaller
value, around 0.7.  Both distributions drop essentially to zero
beyond 1.5 virial velocity.  The shape of these distributions
qualitatively agrees with that obtained by Benson (2005).
Quantitatively, however, both distributions appear slightly
broader than those in Benson (2005). We do not know the exact
reason for the difference, but it may be attributed to the
different ways of selecting the subhalos. For low resolution
simulations, subhalos close to the inner shell $1- \Delta r$ may
be difficult to identify with the friends-of-friends or the
spherical overdensity methods. Benson (2005) attempted to correct
for this effect by considering a minimum radius at which the
subhalo can still be identified using his algorithm.  With the
{\tt SUBFIND} routine (Springel et al. 2001), we are able to
resolve the subhalos even when they are quite close to the host
halos, so we do not need to make this complicated correction.
Instead we simply consider all the subhalos within the spherical
shell between radius $1-\Delta r$ and $1+\Delta r$.

A related quantity to $v_r$ and $v_\theta$ is the infall angle,
defined as \beq \label{eq:theta} \cos \thetaInfall = {v_{\rm r}
\over (v_r^2+v_{\rm \theta}^2)^{1/2}}. \eeq For a radially
infalling subhalo $\thetaInfall=0$. The distribution of
$\thetaInfall$ is shown in the bottom left panel. The infall angle
has a peak around $35^\circ$, and a full-width-at-half-maximum of
about $50^\circ$.

It is well known that halos accrete matter along the large-scale
filaments that connect the halos, this implies that the merger of the
subhalos into the host halos may be anisotropic (Lee, Jing, \& Suto
2005).  Since the halos are generally triaxial (Jing \& Suto 2002),
there could be a correlation between the direction of the subhalo
merger and the shape of the host halo. We therefore determine the
three principal axes for each halo from its inertial tensor within the
virial radius. We define $\theta$ to be the angle between the major
axis of the host halo and the vector from the host halo centre to the
centre of a subhalo, and $\phi$ as the other polar angle ($0\le
\phi<2\pi$). Because of the limited sample size, we will focus on the
spatial distribution as a function of $\mu$ $(\equiv |\cos\theta|)$,
despite of the fact the
distribution also depends on $\phi$ but more weakly than on
$\theta$. The probability distribution of $\mu$,
$df/d\mu$, for the subhalos is shown in the lower right panel of
Fig. 1. If the subhalos merge into the host halo isotropically, then
we expect $df/d\mu$ to be unity.  In contrast to this expectation, $d
f/d\mu$ of the subhalos increases strongly with $\mu$, implying that
subhalos are accreted more preferentially along the major axis of the
host halos.

Since dark halos are triaxial (Jing \& Suto 2002), the dark matter
density at the spherical virial radius is expected to be higher in the
direction of the halo major axis. If the spatial distribution of
subhalos follows that of the dark matter, we would expect a higher $d
f/d\mu$ in the direction of $\mu=1$. The question is whether the increase
of $d f/d\mu$ with $\mu$ can be fully explained by the shape of host
halos.  In the following subsection, we examine this question, i.e.,
consider the phase space distribution in the ellipsoidal coordinate
system in the more realistic triaxial halo model.

\subsection{The phase-space distribution in the triaxial halo model}

For each host halo, we determine the axial ratios $a/c$ and $b/c$ from
their inertia tensor within its virial radius $\rvir$, where $a$,
$b$, and $c$ are the minor, median, and major axes of the halo
respectively. If we rotate the coordinate ($x,y,z$) so that the new
$X$, $Y$ and $Z$ coordinate axes are parallel to the minor, median,
and major axes of the halo, the isodensity surfaces of a halo are
approximately described by (Jing \& Suto 2002)
\begin{equation}
R^2=\frac{c^2 X^2}{a^2}+\frac{c^2 Y^2}{b^2}+Z^2\,.
\end{equation}
We define the boundary of an ellipsoidal halo $\Rvir$ such that the
total mass within $\Rvir$ is equal to the virial mass of the halo
in the spherical model. We then
examine the orbital parameters and density of the subhalos at this
surface. In analogy with the previous subsection, we take a shell of
the upper and lower ellipsoidal radii $(1\pm \Delta R)$ in units
of $\Rvir$, and compute the distributions of the normal and
tangential velocities ($\vn$ and $\vt$)
relative to the ellipsoidal surface as well as the
number density distribution. In this exercise,
we only include the subhalos that have an inward
directed velocity. The infall angle $\thetaInfall$ can be
similarly defined as in eq. (\ref{eq:theta}) with $v_r$ and $v_\theta$ replaced
by $\vn$ and $\vt$.

The results are given in Fig. 2. The distributions of the normal
and tangential velocities are very similar to those in spherical
shells in the previous subsection. The subhalo number density
$n_{\rm tri} (\mu)$, which is the density averaged over $\phi$
directions for a fixed $\mu$ on the surface $\Rvir$, is shown as a
function of $\mu$ in the middle right panel. Notice that in the
triaxial halo model, a flat distribution in $\mu$ implies that the
subhalos follow the idealised ellipsoidal shape. As expected, the
dependence on $\mu$ becomes significantly weaker in the
ellipsoidal shells, but there is still a preference for the
subhalos to merge into their host halos along the major axis.
To facilitate comparisons with previous studies, we also
show in Fig. 2 (bottom two panels) the distributions of the total velocity
$v=({\vn}^2+{\vt}^2)^{1/2}$, and the orbital `circularity'
$\epsilon\equiv J/J_{\rm c}$, where $J$ is angular momentum and
$J_{\rm c}$ is the angular momentum of a circular orbit with the
same energy. The orbital circularity has a broad peak around 0.5,
indicating the infall is in neither purely radial nor tangential
orbits. This result is in good agreement with previous studies
(e.g., Tormen 1997, Fig. 4; Ghigna et al. 1998, Fig. 14).

Next we examine the distribution of the velocities as a function
of the angular positions of the subhalos relative to the major axis of the
ellipsoidal host halo. To do this, we divide the
ellipsoid into three equal bins in $0 \le \mu \le 1$. In Fig. 3,
we show the distributions of the normal
and tangential velocities for these three bins. The distributions
are quite similar. The bottom two panels show
the mean total velocity and the mean infall angle as a function of $\mu$.
Again these two quantities have little dependence on $\mu$.
Therefore we can use the distributions as shown in Figure 2
to more accurately describe the distribution of the subhalo
velocities.

We have also studied the distributions as a function of halo
mass. For this purpose, we increase the lower limit of the host halos
to 5000 particles ($3.1 \times 10^{12}h^{-1}\msun$) which corresponds
to the halo selection parameter set C in Table 1. The results are compared with
those of the host halos with parameter set B, where we included host
halos with more than 500 particles. To increase the number of subhalos
in the case of C, we have adopted $\Delta r=0.2$. As shown in the two
bottom panels of Figure 3, we found the statistical distributions
change very little with reasonable changes of $\Delta r$.  Figure 4
shows the results. No significant difference can be found for the two
velocity distributions and the infall angle distribution. However, the
number density dependence on $\mu$ becomes weaker when we increase the
lower limit of the host halo mass and hence include more subhalos
with smaller $\Msub/\Mhalo$.  If the smaller subhalos follow more
closely the triaxial density profile, then the trend seen in Fig. 4
can be understood because the mass density should follow the triaxial
density model.

To see the preferential accretion of more massive subhalos along the
major axis, in Fig. 5 we divide the subhalos in six bins of
$\Msub/\Mhalo$ (with roughly equal numbers) and show their
distributions of $\mu$. The mass bins are listed in Table 3. The lowest mass-bin subhalos have only a weak
dependence on $\mu$, indicating that they are roughly consistent with
the triaxial shape. However, the highest mass subhalos (top left
panel) with $\Msub/\Mhalo \ga 0.08$ shows dramatic deviation from from
the ellipsoidal accretion, with a strong peak around $\mu=1$ (i.e.,
$\theta=0^\circ$). Fig. 5 clearly illustrates that more massive
subhalos are accreted with a much stronger preference along the major
axis of the host halo. This may have observable consequences for the
satellite galaxies we see today; we return to this important point
briefly in the discussion.

We have shown that the infall of subhalos are
preferentially along the major axis of the parent halo determined
by the moment of inertia. An alternative way of defining the
orientation of a halo is using its spin axis, which can be defined
by the total angular momentum of the particles within the virial
radius. Aubert, Pichon \& Colombi (2004) found that the infall
takes place preferentially in the plane perpendicular to the
direction defined by the spin of the halo. Fig. \ref{fig:spin}
shows the distribution of the angle between the spin axis and the
line connecting the centres of the parent halo and the satellite
in our simulation. Clearly there is a (weak) preference for
satellites to be perpendicular to the spin axis, in agreement with
the findings of Aubert et al. (2004, see their Fig. 8, which is
based on a more sophisticated spherical harmonics analysis). This
is to be expected as the spin axis is statistically perpendicular
to the major axis of the halo, as shown by previous studies
(e.g. Warren et al. 1992, Dubinski 1992).

Many analytical models of the subhalo population require
an accurate knowledge of the initial conditions for the subhalos.
In the following, we provide empirical fitting formulae found
in our numerical simulations.
To facilitate comparisons with Benson (2005), we adopt
the same functional forms to fit the velocity distribution, specifically,
\begin{equation} \label{eq:vr-vtheta}
f(v_n,v_t)=a_1v_t\exp[-a_2(v_t-a_9)^2-b_1(v_t)\{v_n -b_2(v_t)\}^2]
\end{equation}
where
\begin{equation}
b_1(v_t)=a_3\exp[-a_4(v_t-a_5)^2],
\end{equation}
\begin{equation}
b_2(v_t)=a_6\exp[-a_7(v_t-a_8)^2].
\end{equation}
The fitted curves are shown in Fig. 2, and the
fit parameters are listed in Table
\ref{table:fit}. We also use the following function
\begin{equation} \label{eq:mu}
df/d\mu=p_0+p_1\exp(p_2\,\mu^2), ~~\mu \equiv \cos\theta,  0\le \mu \le 1
\end{equation}
to describe the angular distribution of the subhalo accretion. The
function has two parameters, $p_1$ and $p_2$, and $p_0$ is chosen
such that the function is properly normalised when $\mu$ is
integrated from 0 to unity. The fit parameters are listed in Table
\ref{table:dfdmu}. As can be seen, the fitting functions match our
simulation results quite well.

\section{Discussions}

We have used a high-resolution simulation to study the initial
conditions of subhalos when they merge into their host halos. Most of
our results are in good agreement with Benson (2005). One finding of
our study is that massive subhalos are accreted more preferentially
along the major axis of the host halos than the less massive ones.
Our subhalos are identified at present day ($z=0$), but the same trend for massive
subhalos should apply at high redshifts (Kravtsov et al. 2004).  If
the more massive subhalos house satellite galaxies and they survive
until the present day, then the satellites should show a more planar
geometry along the major axis. Interestingly, the satellite galaxies
in the Milky Way appear to lie in a great disk (Kroupa et al. 2005;
see also Willman et al. 2004), almost perpendicular to the stellar
disk. If the major axis of the dark matter halo is perpendicular to
the stellar disk in the Milky Way, then such a distribution, while
puzzling at first, is naturally expected in the CDM (Kang et al. 2005;
Libeskind et al. 2005; Zentner et al. 2005b).

Our result that more massive subhalos are more preferentially
accreted along the major axis of the host halo can be understood
in the cosmic web theory (Bond, Kofman, \& Pogosyan 1996; Lee,
Jing, \& Suto 2005). In this theory, the coherence of the initial
tidal field forms one dimensional filaments. The halos are bridged
by the filaments, and the merger of halos is expected to occur
along the filaments. If the major axis of host halos is determined
by the neighboring massive filaments, we would expect subhalos to
merge preferentially along the major axis. A quantitative
computation is possible for the distribution of $d f/d\mu$
following Lee et al. (2005) which will be explored in a future
work.

The initial conditions we found can be used in analytical
formalisms (e.g., Oguri \& Lee 2004; Taylor \& Babul 2004) to
predict the subsequent evolutions of subhalos and compare these
with observations. Our results make it straightforward to generate
Monte Carlo realizations of the subhalo population. The extended
Press \& Schechter formalism accurately predict the mass function
of the subhalos. For a given halo, the triaxial halo shape can be
sampled using Monte Carlo with the fitting formulae given by Jing
\& Suto (2002). The velocities of the halo can then be obtained
using eq. (\ref{eq:vr-vtheta}). To describe the angular position
of the satellite, the polar angles $(\theta, \phi)$ are needed. To
the first order, the polar angle ($\phi$) is approximately
uniformly distributed between 0 to $2\pi$.  The angle $\theta$ can
be generated using eq. (\ref{eq:mu}). The mass dependence of the
angle $\theta$ can be properly taken into account by choosing
different fit parameters (in Table 3) depending on the value of
$\Msub/\Mhalo$.

In this paper we have analyzed the phase-space information of
subhalos located in a thin shell close to the virial radius.  In
practice, we found that about 30\% of subhalos within the shell
have outgoing velocities. Some of these subhalos and perhaps even
some inward moving subhalos could have been inside the host halo
already, thus some of these subhalos could have evolved
dynamically within the host halo and their phase-space information
cannot be regarded as ``initial'' conditions.  The number of
secondary infall halos can be crudely estimated as follows. Let us
approximate the parent halo as spherical, and assume that there
are $N$ subhalos falling into the parent halo for the first time.
Due to symmetry, we can assume all subhalos move along the
positive $x$-axis, and a fraction $f$ of the subhalos will emerge
as outgoing subhalos on the opposite side of the halo. Presumably
all these $fN$ subhalos will be accreted again in a second infall,
and some fraction (presumably $\la f$) of these subhalos will
emerge as outgoing ones on the positive $x$ side. So in total
there are $N + fN$ infall subhalos, and $fN + f^2N$ outgoing
subhalos. The ratio of infalling and outgoing subhalos is
therefore $(N+fN)/(fN + f^2N) \approx 7/3$, implying $f\sim 3/7$.
Hence the fraction of the secondary infall subhalos is about
$fN/(N+fN) \la 30\%$. Gill et al. (2005) investigated satellite
galaxies in the outerskirts of galaxy clusters from their
high-resolution simulations (see also Gill et al. 2004a,b). They
found that approximately one half of the galaxies with current
cluster-centric distance in the interval 1-2 virial radii of the
host are `backsplash' galaxies that once penetrated the cluster
potential. If one half of these `backsplash' galaxies have
negative radial velocities, then this suggests that the fraction
of subhaloes undergoing secondary infall is about $\sim 25\%$,
consistent with our estimate above.

 To more properly account for this population of
evolved subhalos in detail, one must track the subhalos as a function of
redshift.  This will also be useful for studying the time evolution
of initial conditions. Another issue not studied in detail here is
the $\phi$ dependence of the subhalo spatial distribution. This
quantity is taken to be uniform between 0 to $2\pi$, while
approximately correct to the first order, a more realistic
treatment is desirable. We plan to return to these issues in
future works.

\section*{Acknowledgments}

We thank the referee for a helpful report which improved
the paper. The research is supported by NKBRSF (G19990754), NSFC
(Nos. 10125314, 10373012), and Shanghai Key Projects in Basic
research (No. 04jc14079). SM acknowledges the financial support
of Chinese Academy of Sciences and the
European Community's Sixth Framework Marie
Curie Research Training Network Programme, Contract No.
MRTN-CT-2004-505183 ``ANGLES". HYW and SM wish to
thank the hospitality of Shanghai Astronomical
Observatory during several scientific visits.

\newpage

\begin{table}
\begin{center}
\caption{Halo and subhalo selection parameters: $\Nhalo$ and
$\Nsub$ are the lower limits of the particle number for the host
halos and subhalos respectively. The subhalos are selected within
a spherical shell from $1-\Delta r$ to $1+\Delta r$ in units of
the virial radius. The numbers of selected subhalos are
listed in the last row.}
 \begin{tabular}{cccc}
\\
   \hline\hline
   % after \\: \hline or \cline{col1-col2} \cline{col3-col4} ...
     Model & A & B & C\\
   \hline
   $\Nhalo$ &500 & 500 & 5000 \\
   $\Nsub$ & 20 & 20 & 20 \\
   $\Delta r$ & 0.1 & 0.2 & 0.2 \\
   number of subhalos & 2606 & 4437 & 3470 \\
   \hline
\label{table:haloSelection}
 \end{tabular}
\end{center}
\end{table}

\begin{table}
\begin{center}
\caption{Fit parameters for the velocity distributions
  as defined in eq. (\ref{eq:vr-vtheta}).}
 \begin{tabular}{crr}
\\
   \hline\hline
   % after \\: \hline or \cline{col1-col2} \cline{col3-col4} ...
    parameter & A & B \\
   \hline
   $a_1$ & -- & --  \\
   $a_2$ & 2.12 & 2.62  \\
   $a_3$ &  2.90 & 4.48  \\
   $a_4$ & $-$0.333 & $-$0.525  \\
   $a_5$ & $-$0.490 & 0.238  \\
   $a_6$ & 1.17 & 1.20  \\
   $a_7$ & 0.155 & 0.140  \\
   $a_8$ & $-$0.564 & $-$0.731  \\
   $a_9$ & 0.314 & 0.294  \\
   \hline
 \end{tabular}
\label{table:fit}
\end{center}
\end{table}

\begin{table}
\caption{Fitting parameters for the density distribution $n_{\rm
tri}
  (\mu)$ in different mass bins as defined in eq. (\ref{eq:mu}).  Each
  bin is specified by the lower $M_{\rm l}$ and upper $M_{\rm u}$
  limits of the subhalo mass in units of the mass of the host halo.  }
\begin{center}
\begin{tabular}{cccccccc}
  \hline\hline
  % after \\: \hline or \cline{col1-col2} \cline{col3-col4} ...
   & all subhalos & bin 1 & bin 2 & bin 3 & bin 4 & bin 5 & bin 6 \\
  \hline
   $M_{\rm u}$ & 1 & 1 & 0.07943 & 0.03946 & 0.02542 & 0.01709 & 0.00834 \\
  $M_{\rm l}$ & 0.0 & 0.07943 & 0.03946 & 0.02542 & 0.01709 & 0.00834 & 0.0 \\
  p0 & 0.644 & 0.289 & 0.259 & 0.361 & 0.377 & -4.21 & -617.8 \\
  p1 & 0.194 & 6.33$\times$10$^{-4}$ & 7.37$\times$10$^{-2}$ & 0.159 & 0.258 & 4.98 & 618.8 \\
  p2 & 1.494 & 9.953 & 4.281 & 2.919 & 2.040 & 0.130 & 2.505$\times$10$^{-4}$ \\
\hline \label{table:dfdmu}
\end{tabular}
\end{center}
\end{table}

\newpage

\begin{figure}
  % Requires \usepackage{graphicx}
\epsscale{0.45} \plotone{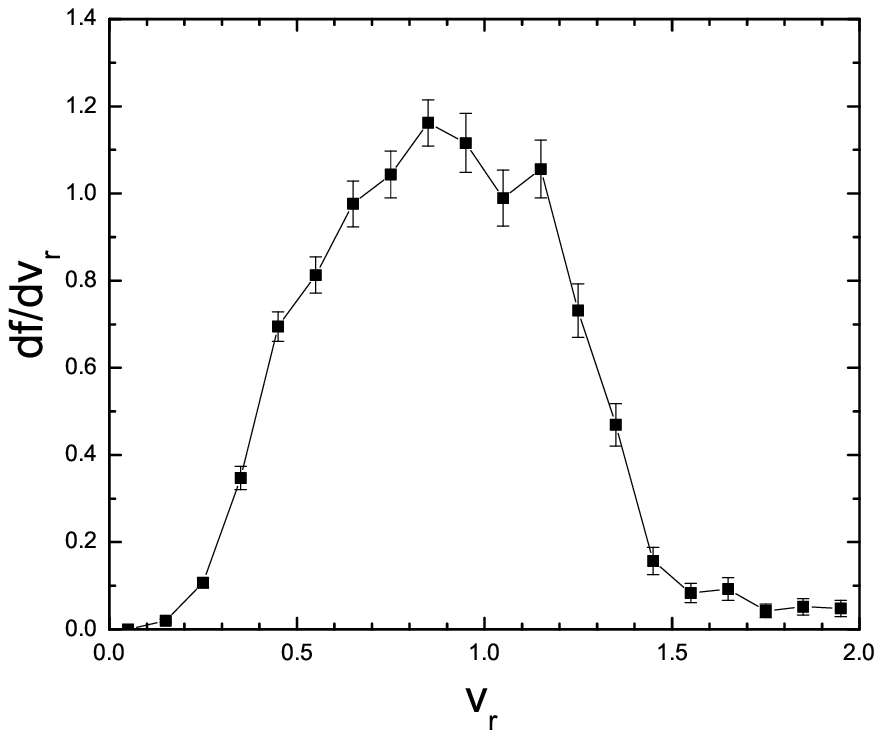}\plotone{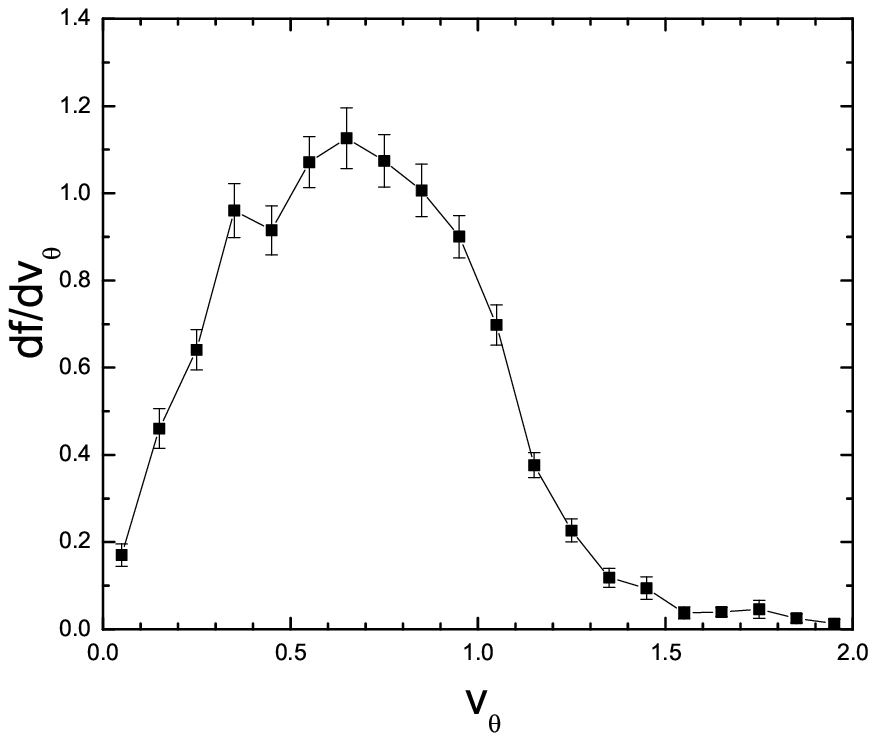}
\plotone{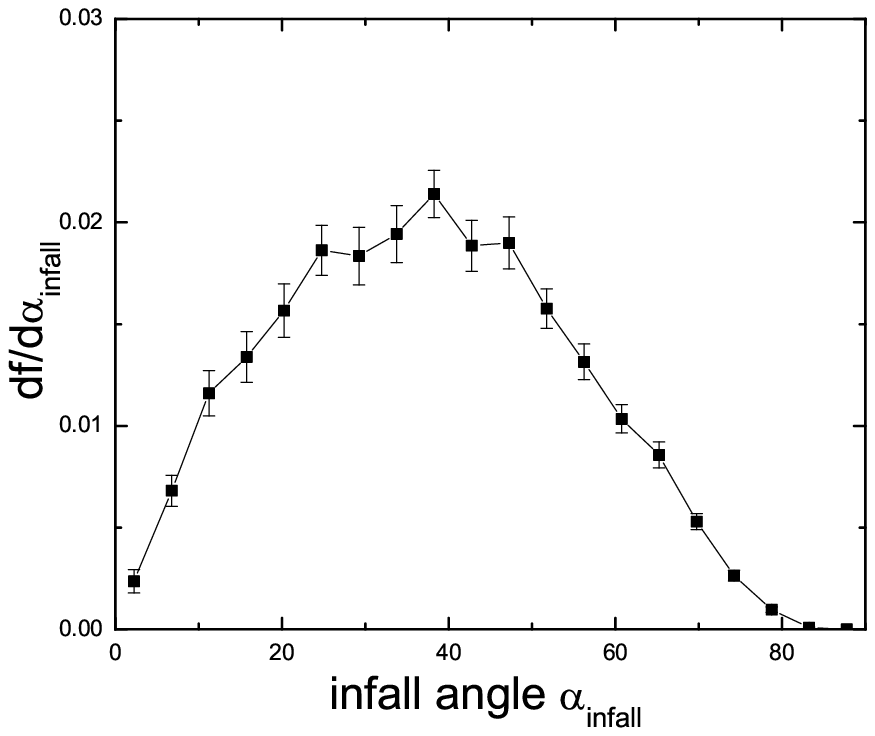}\plotone{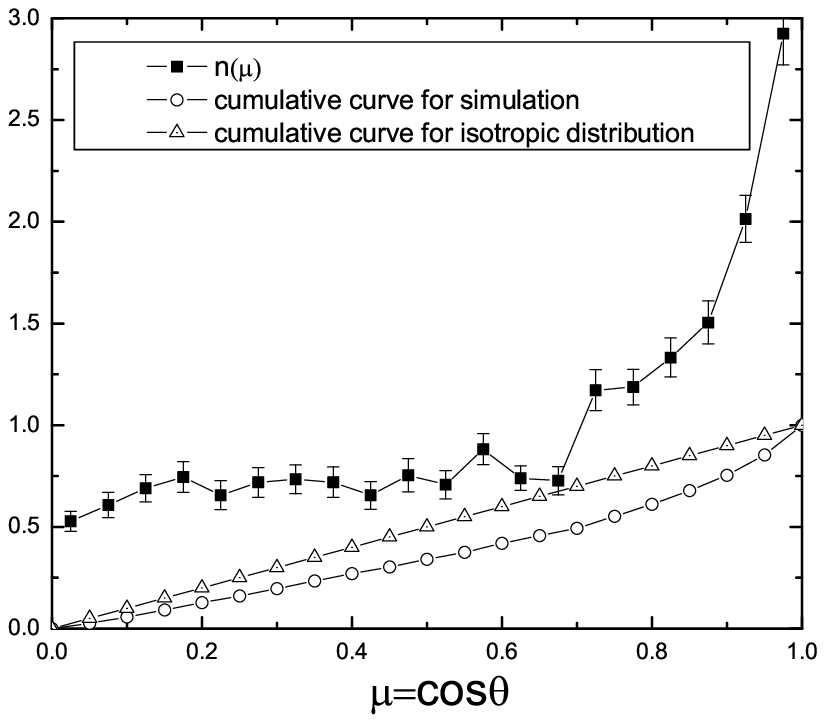}
  \caption{Orbital parameters and spatial distributions for the
 subhalos selected according to the selection parameter set A in Table
 \ref{table:haloSelection} in spherical halo model.  The upper left
 and right panels show the distributions of radial and tangential
 velocities, respectively.  The lower left panel shows the
 distribution of the infall angle, defined in
 eq. (\ref{eq:theta}). In the lower right-hand panel, fill square shows the
 distribution of the number density of subhalos per unit area as a
 function of the angle $\theta$ between the vector from the subhalo
 centre to the host halo centre and the major axis of the host
 halo. For an isotropic population, the distribution $df/d\mu$ is a
 horizontal line with amplitude unity. We also show the
cumulative distributions for the simulation and
an isotropic distribution (open triangles and open circles) in the lower right panel.}
\label{fig:spherical}
\end{figure}

\begin{figure}
  % Requires \usepackage{graphicx}
 \epsscale{0.4} \plotone{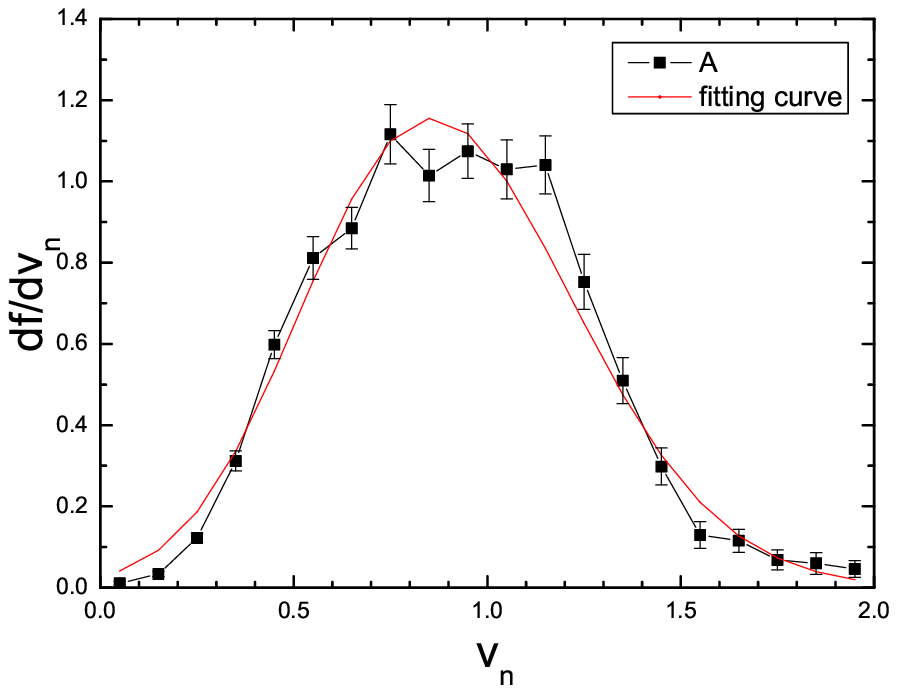}\plotone{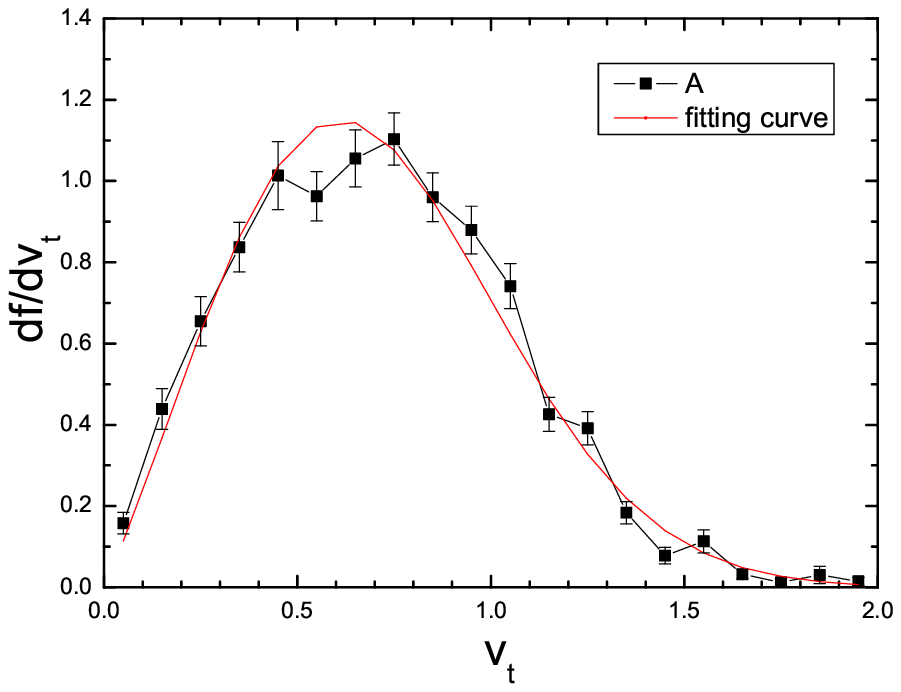}
 \plotone{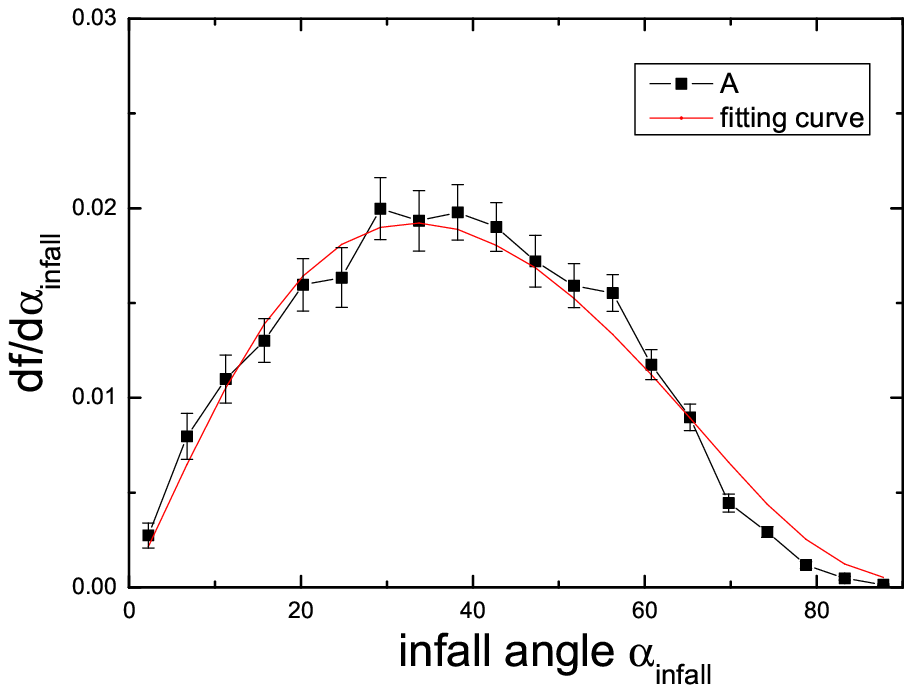}\plotone{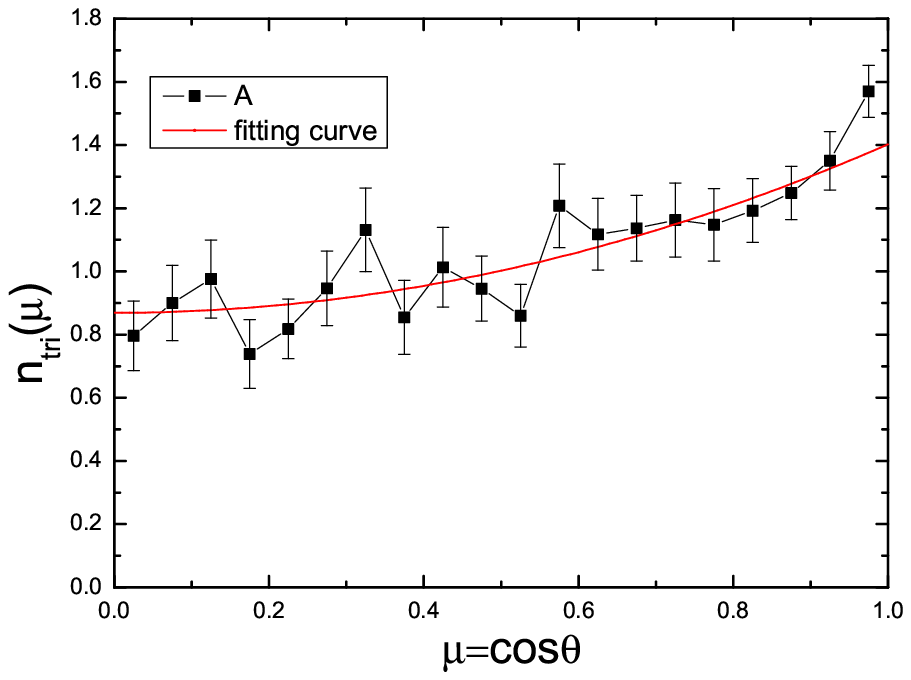}
 \plotone{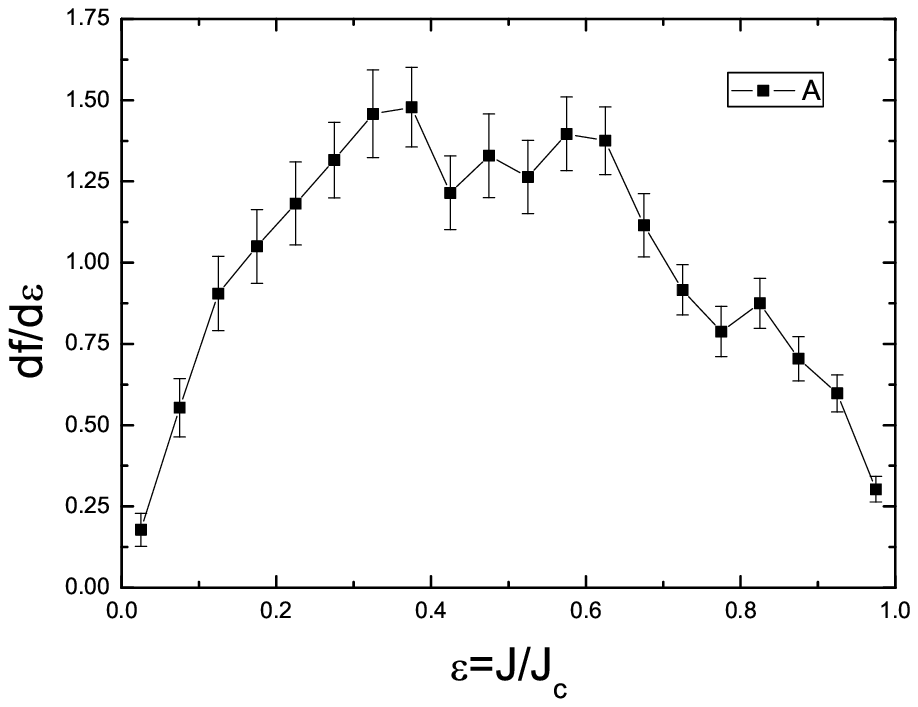}\plotone{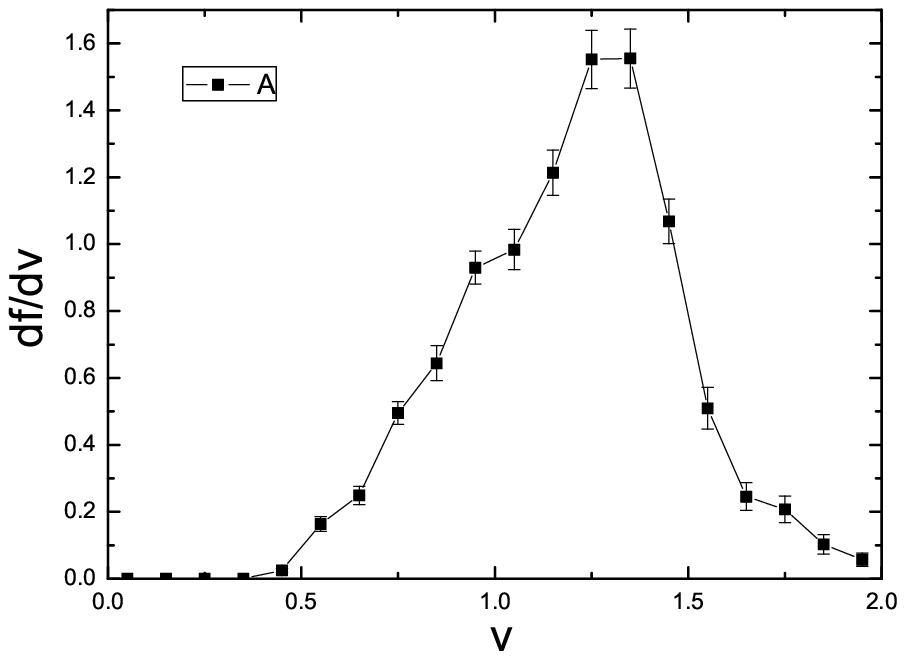}
  \caption{ Orbital parameters and spatial distributions for the
subhalos selected according to the parameter selection set A in
Table \ref{table:haloSelection} in triaxial halo model.  The upper
left and right panels show the distributions of the normal and
tangential velocities respectively.  The middle left panel shows
the distribution of the infall angle.  The middle right panel
shows the number density $n_{\rm tri} (\mu)$ of subhalos at the
surface $\Rvir$ as a function of $\theta$, where $\theta$ is the
angle between the major axis of the host halo and the vector from
the subhalo centre to the host halo centre. The fitting curves as
given in eqs. (\ref{eq:vr-vtheta}) and (\ref{eq:mu}) are shown as
a thin line in each panel. The bottom left and right panels show
the orbital circularity and the total infall velocity
respectively. Notice that the vertical scale in the middle right
panel is different from that in the bottom right panel in Fig. 1.}
\end{figure}

\begin{figure}
  % Requires \usepackage{graphicx}
 \epsscale{0.45} \plotone{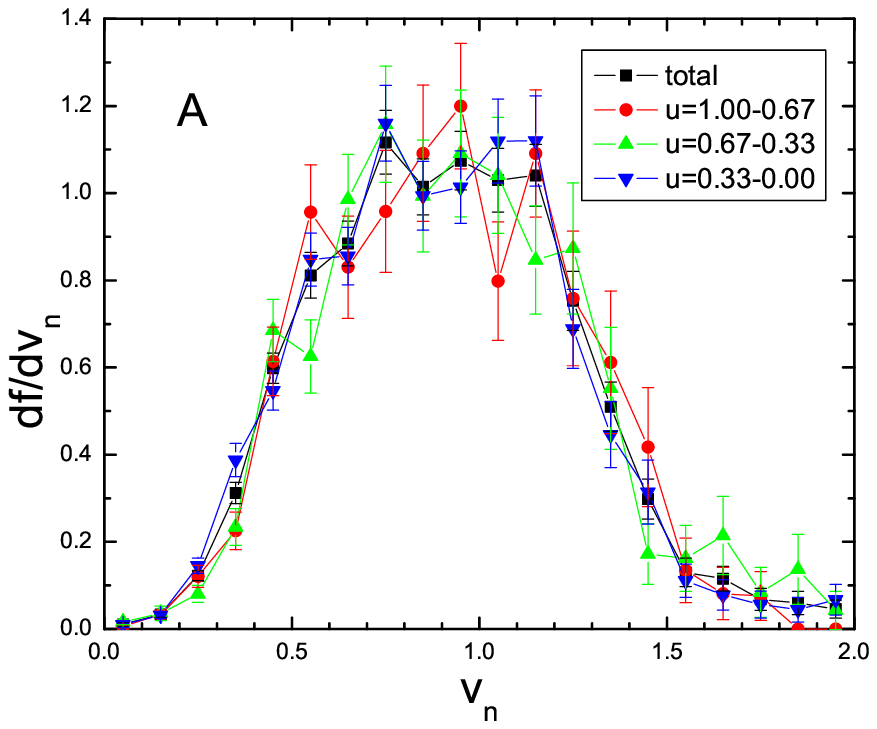}\plotone{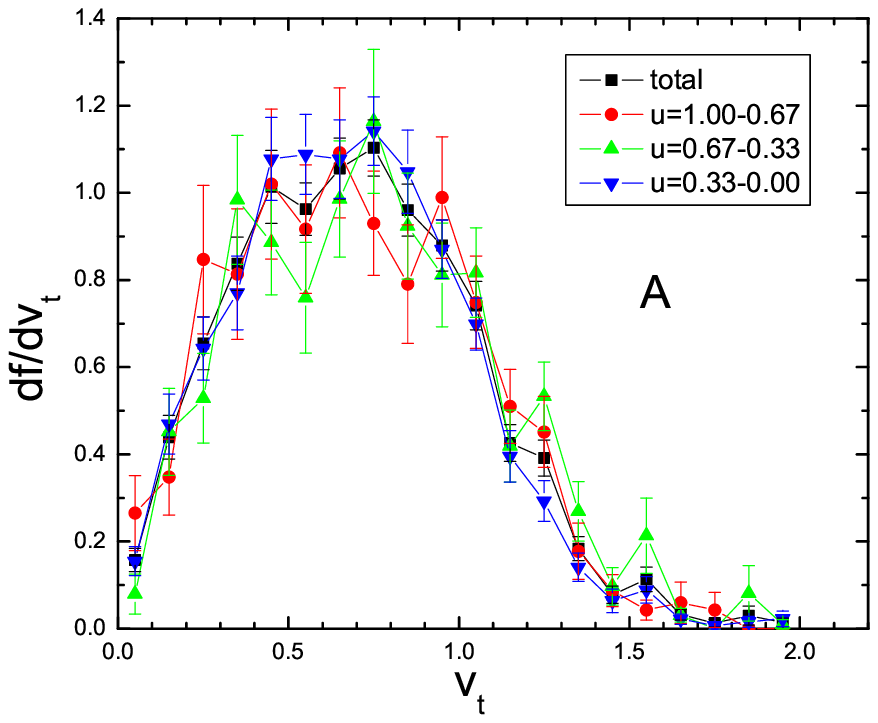}
 \plotone{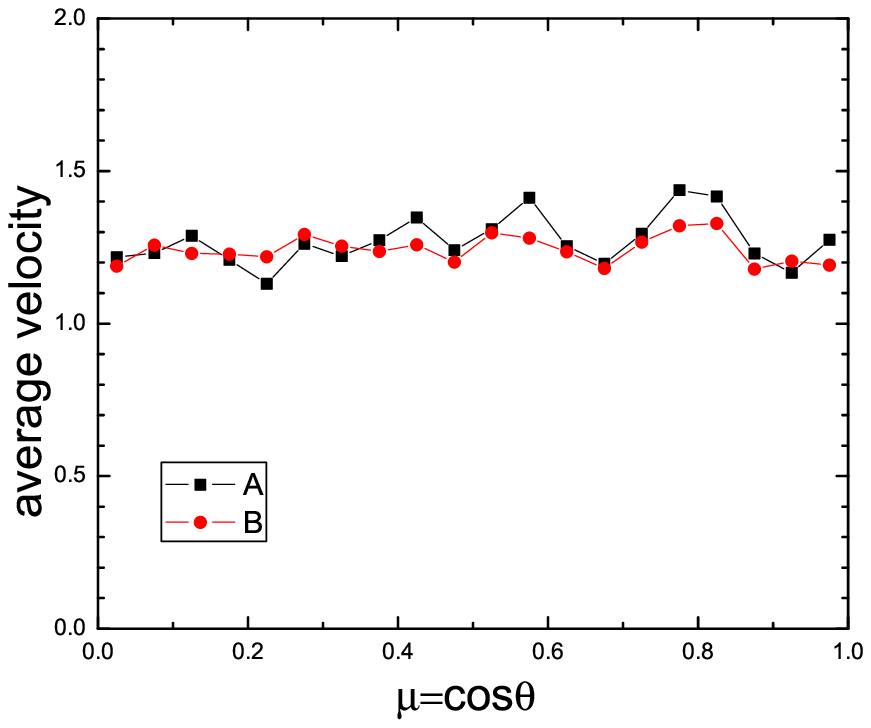}\plotone{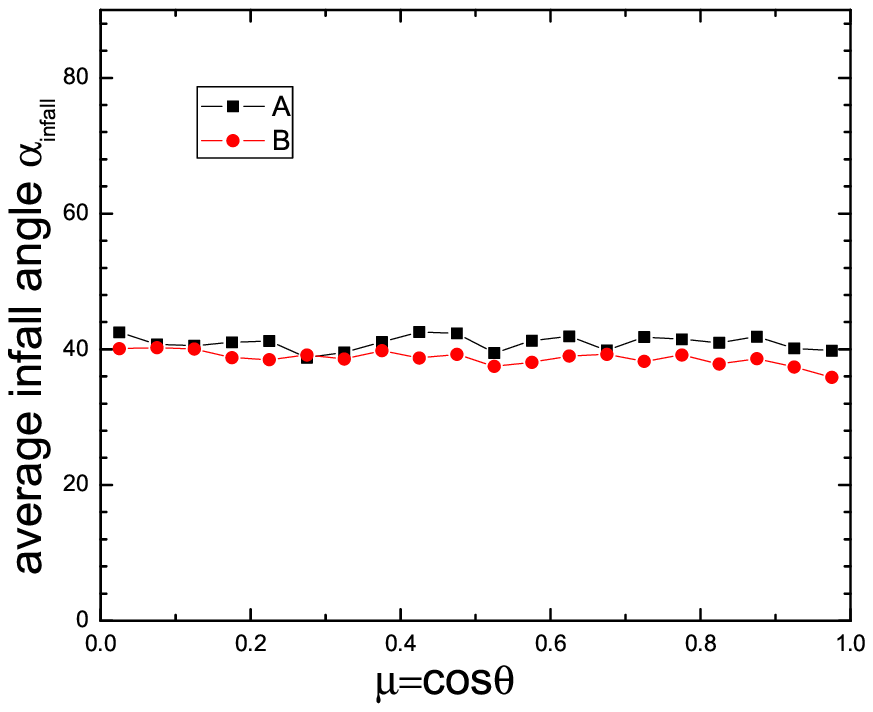}
 \caption{The upper panels plot the distributions of the normal and
 tangential velocities in different ranges of $\mu=\cos\theta$,
 where $\theta$ is the angle between the vector from the subhalo
 centre to the host halos centre and the major axis of the host
 halo. The lower left panel shows the average velocity while the
 lower right panel shows average infall angle as a function of $\mu$.
 }
\end{figure}

\begin{figure}
  % Requires \usepackage{graphicx}
 \epsscale{0.45} \plotone{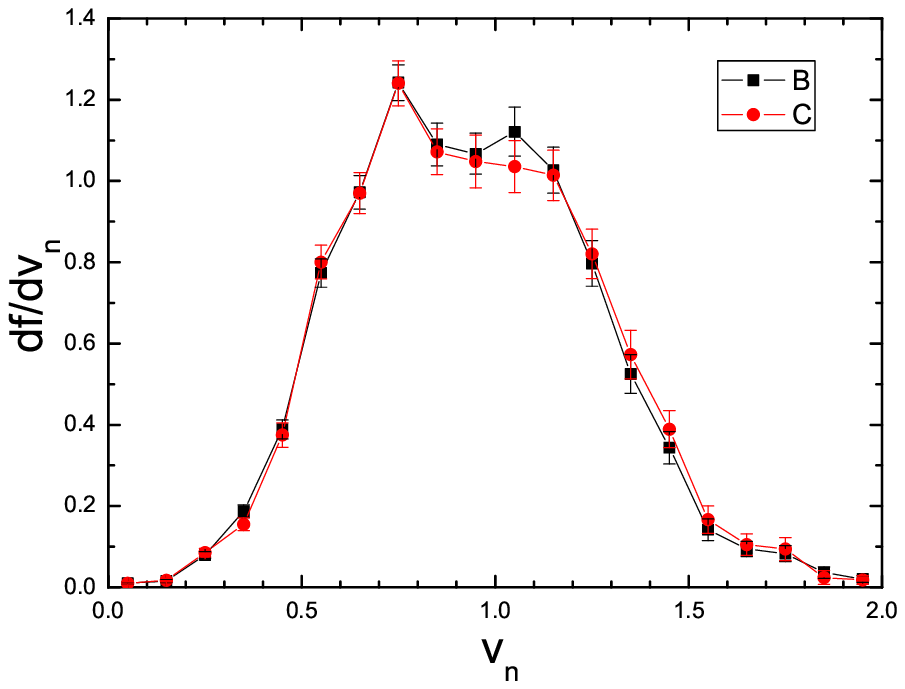}\plotone{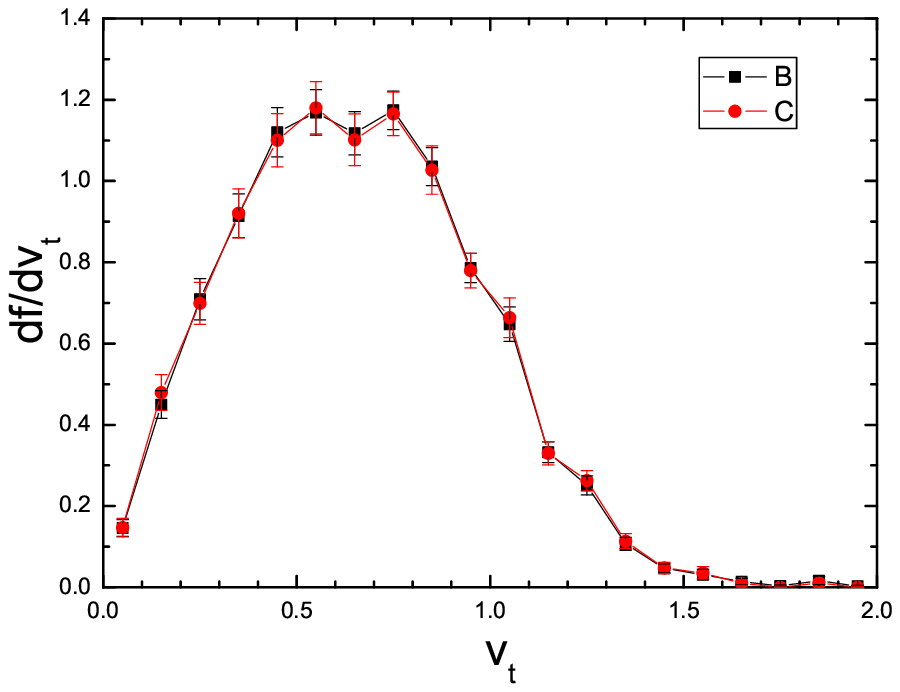}
 \plotone{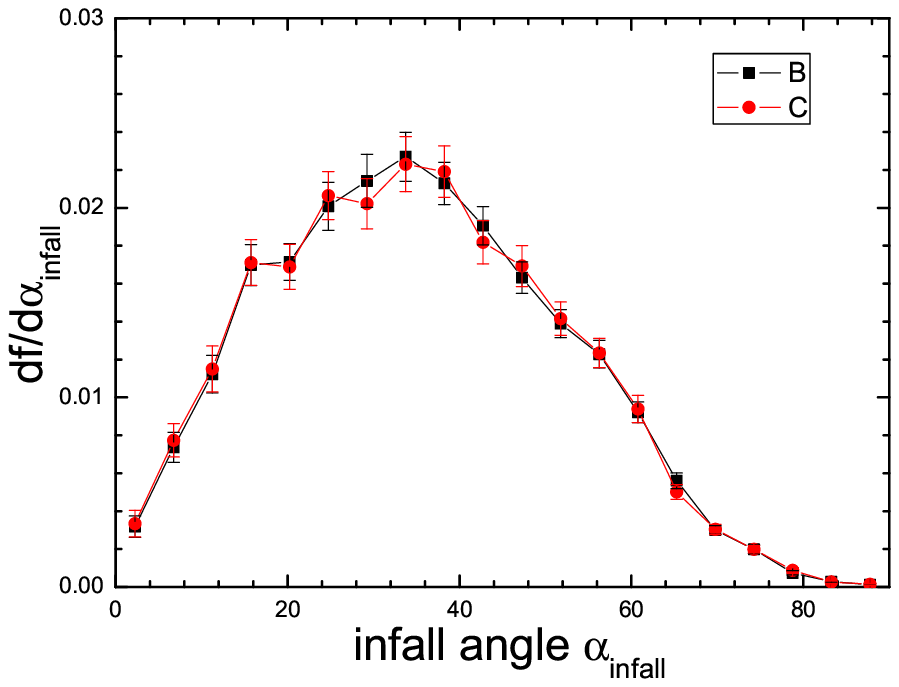}\plotone{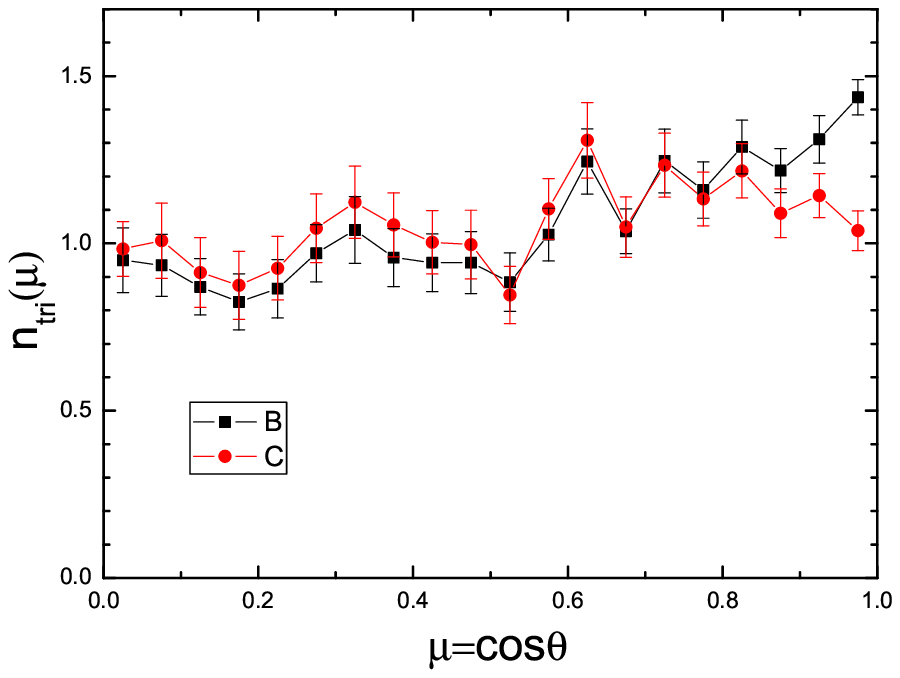}
  \caption{The upper panels show the distributions of the normal and
tangential velocities respectively in the triaxial halo model for
different subhalo masses (Sets B and C). The lower left panel
shows the distribution of the infall angle ($\thetaInfall$) while
the lower right panel shows the number density $n_{\rm tri} (\mu)$
of subhalos at the surface $\Rvir$, where $\mu \equiv \cos\theta$
and $\theta$ is the angle between the major axis of the host halo
and the vector connecting the centres of the subhalo and the host
halo.  The difference between the models B and C in the
bottom right panel is due to the different lower limits of the host
halo mass (see \S 3.2).}
\end{figure}

\begin{figure}
  % Requires \usepackage{graphicx}
 \epsscale{0.4}\plotone{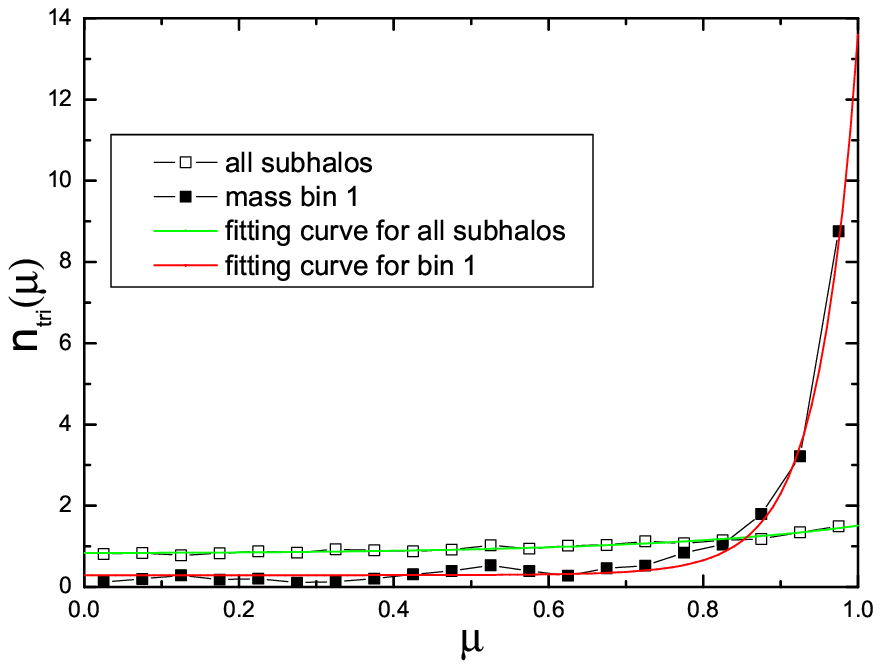}\plotone{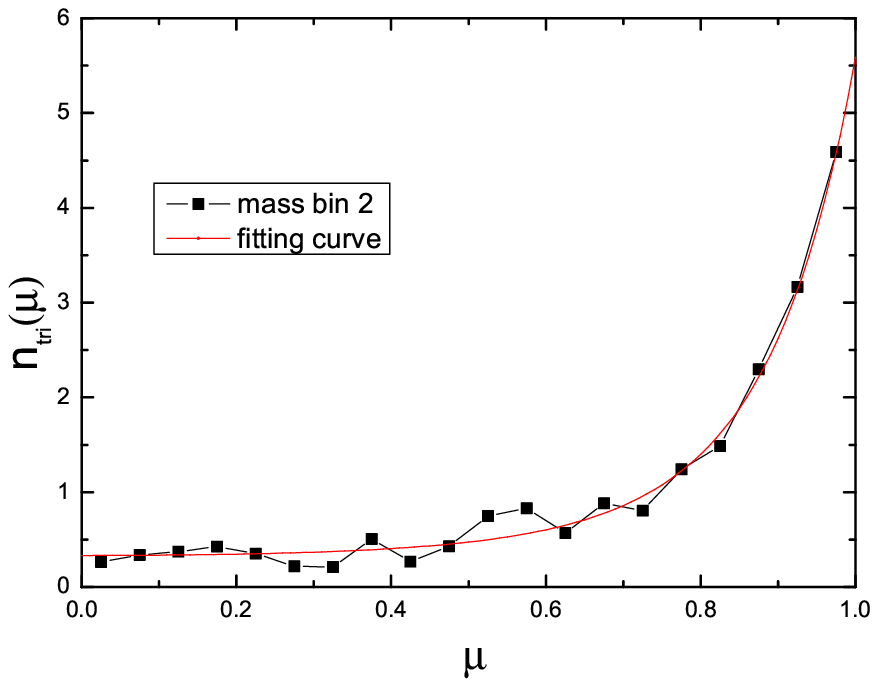}
\plotone{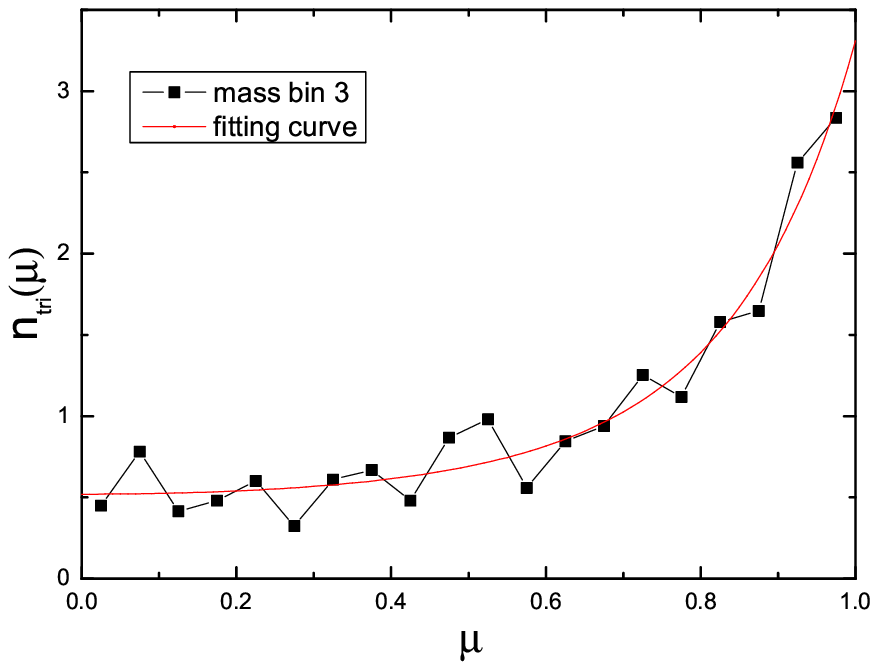}\plotone{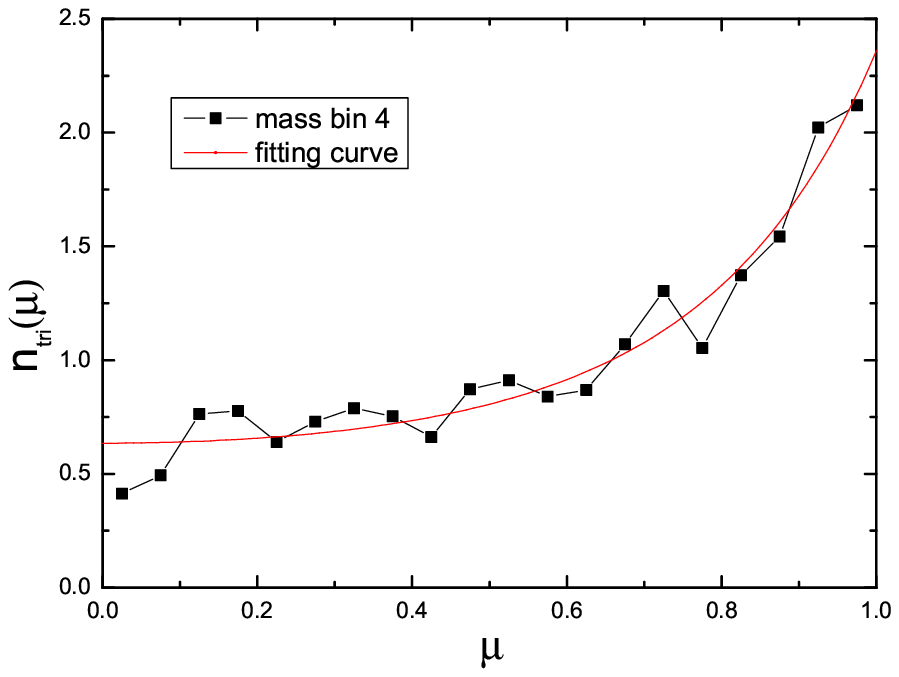}
\plotone{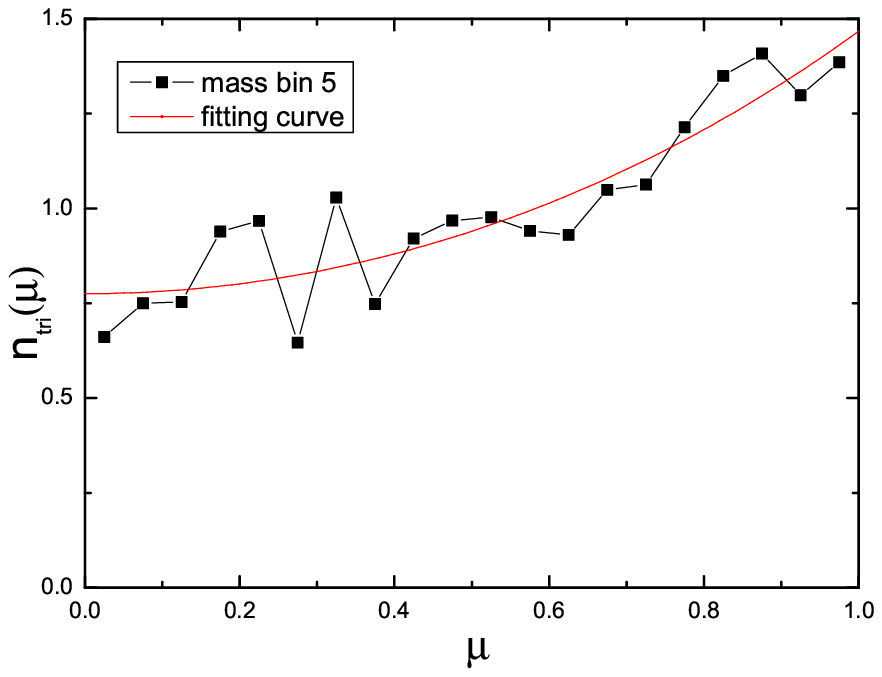}\plotone{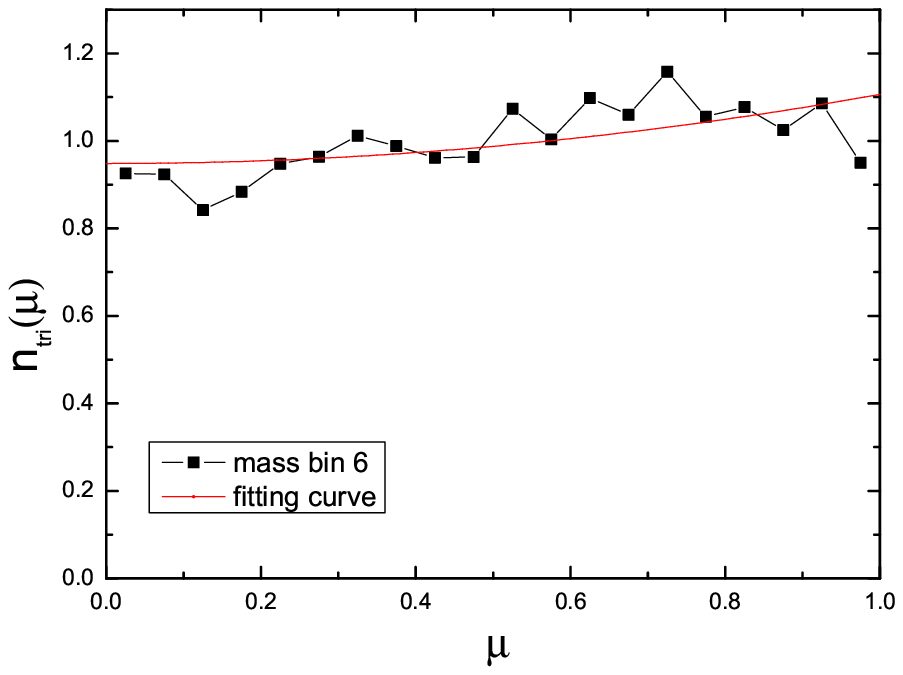}
 \caption{The number density $n_{\rm tri} (\mu)$ of subhalos at the
surface $\Rvir$ together with the best-fit curves in the triaxial
halo model. The subhalos are divided into six bins of $\Msub/\Mhalo$,
and the result for each bin is presented in one panel.  The mass
range for each bin and the corresponding fit parameters (as
defined in eq. \ref{eq:mu}) are listed in Table \ref{table:dfdmu}.
The distribution for all the subhalos is shown in the top left panel.
Notice that the vertical scales are different in different panels.}
\end{figure}

\begin{figure}
  % Requires \usepackage{graphicx}
 \epsscale{0.8} \plotone{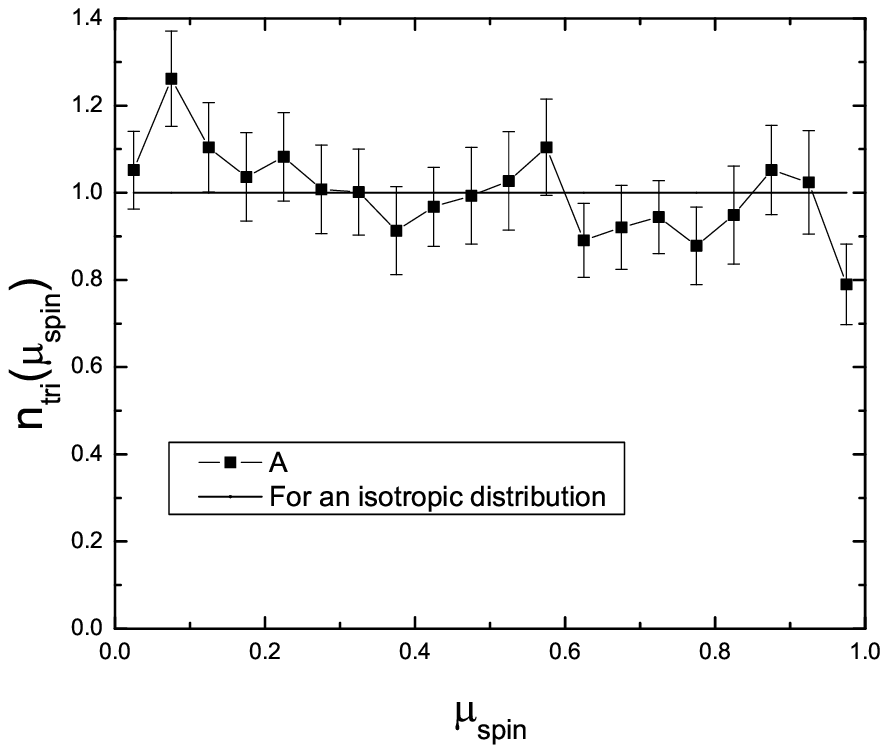}
  \caption{
The distribution of $\mu_{\rm spin}=\cos\thetaspin$ in the
tri-axial model, where $\thetaspin$ is the angle between the
spin axis and the line connecting the centres of the parent and
satellite halos. The straight line shows the prediction for
an isotropic distribution.
\label{fig:spin}
}
\end{figure}


\clearpage

\begin{thebibliography}{}
\bibitem[]{} Aubert D. Pichon C. Colombi S., 2004, MNRAS, 352, 376
\bibitem[]{} Benson A. J., 2005, \mnras, 358, 551
\bibitem[]{} Benson A. J., Lacey C. G., Frenk C. S., Baugh C. M.,
  Cole, S.  \mnras, 2002, 351, 1215
\bibitem[]{} Bertone G., Hooper D.,  Silk J., 2004, Phys. Rep. 405, 279
\bibitem[]{} Bond J., R., Kofman L.,  Pogosyan D., 1996, Nature, 380, 603
\bibitem[]{} Bryan G. L.,  Norman M. L., 1998, ApJ, 495, 80
\bibitem[Col{\'{\i}}n et al.(2000)]{2000ApJ...539..561C} Col{\'{\i}}n P.,
Klypin A.~A.,  Kravtsov A.~V., 2000, \apj, 539, 561
\bibitem[Diemand 04]{diemand04} Diemand J., Moore B.,  Stadel J. 2004, \mnras, 352, 535
\bibitem[Dubinski(1992)]{1992ApJ...401..441D} Dubinski J., 1992, \apj,
401, 441
\bibitem[Gao04a]{ga04a} Gao L., De Lucia G., White S.~D.~M., Jenkins A.,
2004a, \mnras, 352, L1
\bibitem[Gao04b]{gao04b} Gao L., White S.~D.~M., Jenkins A., Stoehr
  F.,  Springel V., 2004b, \mnras, 355, 819
\bibitem[Ghigna et al.(1998)]{1998MNRAS.300..146G} Ghigna S., Moore B.,
Governato F., Lake G., Quinn T.,  Stadel J., 1998, \mnras, 300, 146
\bibitem[Ghigna 00]{ghigna et al. 00} Ghigna S., Moore B., Governato
    F., Lake G., Quinn T., Stadel J., 2000, \apj, 544, 616
\bibitem[Gill et al. (2004a)]{Gil04a} Gill S.~P.~D., Knebe A., Gibson
  B. K. 2004a, \mnras, 351, 399
\bibitem[Gill et al. (2004b)]{Gil04b} Gill S.~P.~D., Knebe A., Gibson
  B. K., Dopita A.~D., 2004b, \mnras, 351, 410
\bibitem[Gill et al. (2005)]{Gil05} Gill S.~P.~D., Knebe A., Gibson
  B. K., 2005, \mnras, 356, 1327
\bibitem[]{} Hockney R. W.,  Eastwood J. W., 1981, Computer
  Simulation Using Particles (New York: McGraw-Hill)
\bibitem[JS02]{js02} Jing Y. P., Suto Y., 2002, ApJ, 574, 538
\bibitem[KJMB]{kjmb04} Kang X., Jing Y. P., Mo H. J.,  B\"orner G.,
  2004, ApJ, in press (astro-ph/0408475)
\bibitem[KMGJ04]{kmgj04} Kang X., Mao S., Gao L., Jing Y. P., 2005,
A\&A, 437, 383
\bibitem[KS96]{ks96} Kitayama T.,  Suto Y., 1996, \apj, 469, 480
\bibitem[Klypin99]{klypin et al. 99} Klypin A., Kravtsov A. V.,
    Valenzuela O., 1999, \apj, 522, 82
\bibitem[Kravtsov et al. 2004]{kravtsov04}Kravtsov A. V., Gnedin O. Y., Klypin A. A., 2004, ApJ, 609, 482
\bibitem[kb03]{kb03} Khochfar S.,  Burkert A., 2004, \mnras, submitted
(astro-ph/0309611)
\bibitem[Kroupa05]{kroupa05} Kroupa P., Theis C., Boily C. M., 2005,
  A\&A, 431, 517
\bibitem[]{} Kochanek C.S.,  Dalal N., 2004, \apj, 610, 69
\bibitem[]{} Lee J., Jing Y.P.,  Suto Y., 2005, ApJ, in press
\bibitem[L05]{} Libeskind N. I., Frenk C. S., Cole S.,  Helly J. C.,
Jenkins A., Navarro J. F., Power C., 2005, astro-ph/0503400
\bibitem[]{} Mao S., 2004, IAU 220, Eds.  M. A. Walker, \&
  K. C. Freeman. ASP 237, p85
\bibitem[]{} Mao S., Schneider P., 1998, \mnras, 295, 587
\bibitem[]{} Mao S., Jing Y. P., Ostriker J. P.,  Weller J. 2004,
  \apj, 604, L5
\bibitem[Moore]{moore et al. 99} Moore B., Ghigna S., Governato
    F. et al., 1999, \apj, 524, L19
\bibitem[OguriLee]{OrugiLee04} Oguri M.,  Lee J., 2004, \mnras, 355, 120
\bibitem[Okamoto  Habe(1999)]{1999ApJ...516..591O} Okamoto T.,  Habe A., 1999, \apj, 516, 591
\bibitem[]{} Sheth R. K., 2003, MNRAS, 345, 1200
\bibitem[Springel01]{s01} Springel V., White S.~D.~M.,
  Tormen G.,  Kauffmann G., 2001, \mnras, 328, 726
\bibitem[WMAP]{spergel} Spergel D.N.S. et al., 2003, \apjs, 148, 175
\bibitem[Stoehr03]{stoehr03} Stoehr F., White S. D. M., Springel V.,
         Tormen G.,  Yoshida N. 2003, \mnras, 345, 1313
\bibitem[Taylor  Babul(2001)]{2001ApJ...559..716T} Taylor J.~E., Babul A., 2001, \apj, 559, 716
\bibitem[tb04]{TaylorBabul2004} Taylor J. E.,  Babul A., 2004, \mnras, 348, 811
\bibitem[ts03]{TaylorSilk2003}  Taylor J. E.,  Silk J., 2003, \mnras,  339, 505
\bibitem[torman97]{Tormen 1997} Tormen G., 1997, \mnras, 290, 411
\bibitem[vit02]{vit02} Vitvitska M., Anatoly A., Kravtsov A. V.,
  Wechsler R. H., Primack J. R.,  Bullock J. S., 2002, \apj, 581, 799
\bibitem[Warren et al.(1992)]{1992ApJ...399..405W}
Warren M.~S., Quinn P.~J., Salmon J.~K.,  Zurek W.~H., 1992, \apj, 399, 405
\bibitem[Willman]{willman} Willman B., Governato F., Dalcanton J. J.,
  Reed D.,  Quinn T. 2004, \mnras, 353, 639
\bibitem[]{} van den Bosch F. C., Lewis G. F., Lake G., Stadel J.,  1999, 515, 50
\bibitem[]{} van den Bosch F. C., Tormen G., Giocoli C., 2005, MNRAS, 359, 1029
\bibitem[]{} White S. D. M.,  Rees M. J., 1978, \mnras, 183, 341
\bibitem[]{} Zentner A. R., Bullock J. S., 2003, 598, 49
\bibitem[]{} Zentner A. R., Berlind A. A., Bullock J. S., Kravtsov
  A. V., Wechsler, R. H., 2005a, ApJ, 624, 505
\bibitem[z05]{Zentner05} Zentner  A. R., Kravtsov A. V., Gnedin
  O. Y., Klypin A. A. 2005b, astro-ph/0502496
\end{thebibliography}
\end{document}